\documentclass[aps,prc,showpacs,showkeys,twoside,twocolumn,byrevtex,floatfix]{revtex4-2}

\usepackage[dvips]{graphicx} 
\usepackage{color}           
\usepackage{amsmath,amssymb,amsfonts}
\usepackage{bm}
\usepackage{mathtools}
\newcommand{\nc}{\newcommand}           
\nc{\vc}[1]     {\mbox{\boldmath $#1$}} 
\nc{\mapleft}[1]{                       
 \smash{\mathop{                      %
  \hbox to 0.90cm{\rightarrowfill} }\limits_{#1}}}

\nc{\beq}     {\begin{eqnarray}}
\nc{\eeq}    {\end{eqnarray}}
\nc{\bra}       {\langle}               
\nc{\ket}       {\rangle}               
\nc{\bras}[1]   {\langle#1|}            
\nc{\kets}[1]   {|#1\rangle}            
\nc{\del}       {\partial}              

\newcommand{\lw}[1]{\smash{\lower1.75ex\hbox{#1}}}

\nc{\mydraft}	{\setlength{\topmargin}{-1.5cm}}
\mydraft
\begin{document}

\title{
  Variation of multi-Slater determinants in antisymmetrized molecular dynamics
  and its application to $^{10}$Be with various clustering
}

\author{Takayuki Myo}
\email{takayuki.myo@oit.ac.jp}
\affiliation{General Education, Faculty of Engineering, Osaka Institute of Technology, Osaka, Osaka 535-8585, Japan}
\affiliation{Research Center for Nuclear Physics (RCNP), Osaka University, Ibaraki, Osaka 567-0047, Japan}

\author{Mengjiao Lyu}
\affiliation{College of Science, Nanjing University of Aeronautics and Astronautics, Nanjing 210016, China}

\author{Qing Zhao}
\affiliation{School of Science, Huzhou University, Huzhou 313000, Zhejiang, China}

\author{Masahiro Isaka}
\affiliation{Hosei University, Chiyoda-ku, Tokyo 102-8160, Japan}

\author{Niu Wan}
\affiliation{School of Physics and Optoelectronics, South China University of Technology, Guangzhou 510641, China}

\author{Hiroki Takemoto}
\affiliation{Faculty of Pharmacy, Osaka Medical and Pharmaceutical University, Takatsuki, Osaka 569-1094, Japan} 
\affiliation{Research Center for Nuclear Physics (RCNP), Osaka University, Ibaraki, Osaka 567-0047, Japan}

\author{Hisashi Horiuchi}
\affiliation{Research Center for Nuclear Physics (RCNP), Osaka University, Ibaraki, Osaka 567-0047, Japan}

\date{\today}

\begin{abstract}%
  We propose a method to optimize the multi-Slater determinants of the antisymmetrized molecular dynamics (AMD)
  in the linear combination form and apply it to the neutron-rich $^{10}$Be nucleus.
  The individual Slater determinants and their weights in the superposition are determined simultaneously
  according to the variational principle of the energy of the total wave function.
  The multi-AMD basis states of $^{10}$Be show various cluster structures as well as the shell-model type.
  In the cluster configurations, different intercluster distances are superposed automatically indicating the role of the generator coordinates.
  We further introduce a procedure to obtain the configurations for the excited states imposing the orthogonal condition to the ground-state configurations.
  In the excited states of $^{10}$Be, the linear-chain-like structure is confirmed consisting of various clusters.
  The energy spectrum using the obtained basis states reproduces the experiments.
  The present framework can be the method to find the optimal multi-configuration for nuclear ground and excited states.
\end{abstract}

\maketitle

\section{Introduction}

Nuclear clustering is one of the important properties of nuclei as well as the mean-field states \cite{ikeda68,oertzen06,horiuchi12,freer18,zhou20},
in which some nucleons form a cluster such as an $\alpha$ particle, and are spatially developed in nuclei.
A typical case is a $^8$Be nucleus, which decays into two $\alpha$ particles.
The Hoyle state, the $0^+_2$ state of $^{12}$C, is considered to have a 3$\alpha$ structure
and this state is the resonance located near the threshold energy of the 3$\alpha$ breakup.

The clustering phenomena of nuclei have a variety as indicated by the Ikeda diagram \cite{ikeda68}.
Theoretically, it is important to investigate the possible cluster states as well as the mean-field states without any assumption as much as possible.
For this purpose, the antisymmetrized molecular dynamics (AMD) is one of the possible models of nuclear many-body wave functions in the Slater determinant form
\cite{kanada03,kimura16}.
In AMD, the nucleon wave function has a Gaussian wave packet with a specific centroid position in phase space.
The distribution of the centroid parameters controls the appearance of the cluster and mean-field states of nuclei
and these parameters are determined according to the variational principle of the total energy.

The AMD wave function can be extended to the multi-configuration (multi-Slater determinants) by applying the generator coordinate method (GCM),
in which various sets of the Gaussian parameters are used in each of the Slater determinants.
The weights of the Slater determinants are determined in the energy minimization of the total system.
In AMD with GCM, the basis states are usually produced by imposing the constraints on the AMD wave functions, such as deformation and radius,
and one often regards the lowest-energy state in the specific constraint as one of the basis states in GCM.
However, it is not trivial whether the basis states obtained with the constraints can be the optimal configurations to be superposed in the results of GCM.

In the mean-field model of nuclei, the method to superpose the optimal Slater determinants has been discussed
in the multi-configuration Hartree-Fock theory \cite{faessler71,kim71}
and also in the self-consistent multiparticle-multihole configuration mixing \cite{pillet17}.

Recently, we developed two variational methods for nuclei using the AMD wave function to treat the short-range and tensor correlations induced by the realistic nucleon-nucleon interaction.
One is the ``tensor--optimized antisymmetrized molecular dynamics'' (TOAMD) \cite{myo15,myo17a,lyu20,myo22},
in which we introduce the correlation functions of the central-operator and tensor-operator types.
The other is ``high--momentum antisymmetrized molecular dynamics'' (HM-AMD) \cite{myo17e,lyu18,zhao21,isaka22}, 
in which high-momentum components are treated using the complex Gaussian centroids.
Two methods are also applied to the nuclear matter calculations \cite{wan22,yamada18}.
In these methods, when many basis states are superposed with various variational parameters, the solutions can show convergence,
but the calculation cost becomes high. We need a scheme to find the optimal basis states in these variational methods.

In this paper, we propose a method to search for the optimal multi-Slater determinants using the AMD basis states for nuclei.
In AMD, the Gaussian centroids are variational parameters and
the imaginary-time evolution of them is solved to gain the total energy until getting the converging solutions, which is so-called the cooling equation. 
We extend this cooling method to treat the multi-basis states and
we can determine all the parameters in the multi-basis states and the weights of the basis states simultaneously.

In the method, all the configurations (Slater determinants) are determined in the energy minimization of the total wave function.
This would be useful to find the appropriate configurations of the cluster and mean-field types in the superposition of the basis states.
The present framework would be applicable to the methods of TOAMD and HM-AMD with realistic interactions.
Optimization of the multi-configuration is also available in the neural network approach
and our collaborators have recently applied it to the cluster states of $^{12}$C \cite{cheng23}.

In this paper,
we apply this scheme to a neutron-rich $^{10}$Be nucleus in the description of the ground and excited states including the cluster and shell-like states.
There have been theoretical studies of $^{10}$Be focusing on the clustering structures \cite{kanada99,itagaki00,suhara10,kobayashi11,shikata20,isaka15}.
Using the same Hamiltonian as used in these studies, we discuss $^{10}$Be and compare the results with the previous works.

In Sec.~\ref{sec:method}, we explain the formulation of the variation of the multi-Slater determinants in the AMD wave functions.
In Sec.~\ref{sec:result}, we discuss the results of $^{10}$Be. 
In Sec.~\ref{sec:summary}, we summarize this work.

\section{Theoretical methods}\label{sec:method}

\subsection{Hamiltonian}\label{sec:ham}
We use the Hamiltonian with a two-body nucleon-nucleon interaction for mass number $A$ as
\begin{eqnarray}
    H
&=& \sum_i^{A} t_i - T_{\rm c.m.} + \sum_{i<j}^{A} v_{ij}\, ,
    \label{eq:Ham}
    \\
    v_{ij}
&=& v_{ij}^{\rm central} + v_{ij}^{LS} + v_{ij}^{\rm Coulomb}.
\end{eqnarray}
Here, $t_i$ and $T_{\rm c.m.}$ are the kinetic energies of each nucleon and the center-of-mass, respectively.
Following the previous works of $^{10}$Be in similar models \cite{itagaki00,suhara10,kobayashi11,shikata20},
we use the effective nucleon-nucleon interaction $v_{ij}$ of Volkov No.2 central force with Majorana parameter $M=0.6$
and the Bartlett and the Heisenberg parameters $B=H=0.125$, 
G3RS LS force with the strength of 1600 MeV, and the point Coulomb force for protons.
This Hamiltonian is the same as that used in other calculations \cite{itagaki00,suhara10,kobayashi11,shikata20}
except for the LS strength of 2000 MeV \cite{itagaki00}.

In the present Hamiltonian, 
the total energy of the $\alpha$ particle is $-27.6$ MeV with the $(0s)^4$ configuration and
the $\alpha+\alpha+n+n$ threshold energy of $^{10}$Be is $-55.2$ MeV,
which is close to the experimental value of $-56.6$ MeV.

\subsection{Antisymmetrized molecular dynamics (AMD)}\label{sec:AMD}

We explain the framework of AMD for nuclear many-body systems.
The AMD wave function $\Phi_{\rm AMD}$ is a single Slater determinant of $A$-nucleons given as
\begin{eqnarray}
\Phi_{\rm AMD}
&=& \frac{1}{\sqrt{A!}} {\rm det} \left\{ \prod_{i=1}^A \phi_i(\bm{r}_i) \right\}~,
\label{eq:AMD}
\\
\phi_i(\bm{r})&=&\left(\frac{2\nu}{\pi}\right)^{3/4} e^{-\nu(\bm{r}-\bm{Z}_i)^2} \chi_{\sigma,i} \chi_{\tau,i},
\\
\chi_{\sigma,i} &=& \alpha_i \kets{\uparrow} + \beta_i \kets{\downarrow}.
\label{eq:Gauss}
\end{eqnarray}
The single-nucleon wave function $\phi_i(\bm{r})$ has a Gaussian wave packet with a common range parameter $\nu$
and the individual centroid position $\bm{Z}_i$ with an index $i=1,\cdots,A$.
We set $\nu=0.235$ fm$^{-2}$, which is the same value as used in Refs. \cite{itagaki00,suhara10,kobayashi11,shikata20}.
We impose the condition of $\sum_{i=1}^A \bm{Z}_i={\bf 0}$.
The spin part $\chi_{\sigma}$ is a mixed state of the up ($\uparrow$) and down ($\downarrow$) components for the $z$ direction
and the isospin part $\chi_{\tau}$ is a proton (p) or neutron (n).
The variational parameters of each nucleon are $\bm{Z}_i$, $\alpha_i$, and $\beta_i$ being complex numbers.

In AMD, the energy variation is performed by solving the cooling equation,
in which the variational parameters are determined in terms of the imaginary-time evolution of the following equation:
\begin{equation}
  \begin{split}
    X_i &=\{\bm{Z}_i,\alpha_i,\beta_i\},
    \\
    \Phi^\pm_{\rm AMD}&= P^\pm \Phi_{\rm AMD},
    \\
    E^\pm_{\rm AMD} &= \dfrac{ \bra \Phi^\pm_{\rm AMD}|H| \Phi^\pm_{\rm AMD} \ket }{\bra \Phi^\pm_{\rm AMD}| \Phi^\pm_{\rm AMD} \ket},
    \\
    \dfrac{{\rm d} X_i}{{\rm d} t}&= - \mu \dfrac{\partial E^\pm_{\rm AMD}}{ \partial X_i^*},\quad \mbox{and c.c}.
  \end{split}
  \label{eq:cooling}
\end{equation}
Here $X_i$ stands for the set of parameters of a single nucleon and $P^{\pm}$ is a parity-projection operator and $\mu$ is an arbitrary positive number. 
This equation satisfies the condition of the energy minimization as ${\rm d}E^\pm_{\rm AMD}/{\rm d}t \leq 0 $.
In this study, we solve the cooling equation with the parity projected wave function until we get the convergence of $E^\pm_{\rm AMD}$.

Finally, using the angular momentum projection operator $P^J_{MK}$,
one obtains the eigenstate of the angular momentum $J$ with quantum numbers of $M$ and $K$, and the parity ($\pm$),
and also the total energy $E^{J^\pm}_{\rm AMD}$.
\begin{eqnarray}
  \begin{split}
  \Psi^{J^\pm}_{MK,{\rm AMD}}
  &= P^J_{MK}P^{\pm} \Phi_{\rm AMD},
  \\
  E^{J^\pm}_{\rm AMD}
  &= \frac{\langle \Psi^{J^\pm}_{MK,{\rm AMD}} | H | \Psi^{J^\pm}_{MK,{\rm AMD}}\rangle}{\langle\Psi^{J^\pm}_{MK,{\rm AMD}} |\Psi^{J^\pm}_{MK,{\rm AMD}}\rangle}.
  \end{split}
  \label{eq:projection}
\end{eqnarray}
The formulas of the Hamiltonian matrix elements are explained in Refs. \cite{brink66,horiuchi70}.

The AMD wave function is usually extended to the multi-configuration (multi-Slater determinants) applying GCM.
In GCM, the parameter set of $\{X_i\}$ in the single AMD configuration is usually determined by imposing specific constraints in the cooling equation.
The total wave function $\Psi_{\rm GCM}$ is a linear-combination form of the projected AMD basis states denoted simply as $\Psi_n$
with the basis index of $n=1,\cdots,N$.
\begin{eqnarray}
  \begin{split}
   \Psi_{\rm GCM}
&= \sum_{n=1}^N C_n\,  \Psi_n \,,
   \\
   H_{mn}
&= \langle\Psi_{m} | H |\Psi_{n}\rangle \,,
   \quad
   N_{mn}
=  \langle\Psi_{m} | \Psi_{n} \rangle \,,
   \\
   E_{\rm GCM} &= \dfrac{ \bra \Psi_{\rm GCM} |H| \Psi_{\rm GCM} \ket }{\bra \Psi_{\rm GCM}| \Psi_{\rm GCM} \ket}, 
  \end{split}
   \label{eq:linear}
\end{eqnarray}
where the labels $m$ and $n$ represent the set of the quantum numbers of the projected AMD basis states
in Eq.~(\ref{eq:projection}) with $\{X_i\}$. These basis states are non-orthogonal to each other.
The Hamiltonian and norm matrix elements are given as $H_{mn}$ and $N_{mn}$, respectively.
From the variational principle of the total energy, $\delta E_{\rm GCM}=0$,
one solves the following eigenvalue problem and obtains the total energy $E_{\rm GCM}$ and the coefficients $\{C_n\}$ for each $J^\pm$.
\begin{eqnarray}
   \sum_{n=1}^N \bigl( H_{mn} - E_{\rm GCM} N_{mn} \bigr) C_n &=& 0.
   \label{eq:eigen}
\end{eqnarray}
\subsection{Variation of multi-Slater determinants}\label{sec:multi}

We propose the method to optimize the multi-AMD basis states simultaneously without any constraint in the energy variation of the total system,
which is the extension of the cooling equation in Eq.~(\ref{eq:cooling}).
In this study, in the energy variation to construct the basis states, every wave function and equation is given with the parity projection
and hereafter we simply write them omitting the notation of parity $(\pm)$.
We express the total wave function $\Phi$ in the linear combination of the basis states $\{\Phi_n\}$
with parity projection in the intrinsic frame.
Each of the AMD configurations has variational parameters of $X_{n,i}=\{\bm{Z}_{n,i},\alpha_{n,i},\beta_{n,i},C_n\}$ including the coefficient $C_n$.
The total energy $E_{\rm GCM}$ and the cooling equation to minimize $E_{\rm GCM}$ are expressed as 
\begin{equation}
  \begin{split}
   \Phi &= \sum_{n=1}^N C_n\,  \Phi_n , \qquad
   E_{\rm GCM} = \dfrac{ \bra \Phi |H| \Phi \ket }{\bra \Phi | \Phi \ket},
   \\
    \dfrac{{\rm d} X_{n,i}}{{\rm d} t}&= - \mu \dfrac{\partial E_{\rm GCM}}{ \partial X_{n,i}^*},\quad \mbox{and c.c}.
  \end{split}
  \label{eq:multi}
\end{equation}
Solving these equations numerically, one can obtain the total energy $E_{\rm GCM}$
and optimize all the parameters in the multi AMD basis states simultaneously. 
Finally, after the angular momentum projection of the basis states $\{\Phi_n\}$, one solves the eigenvalue problem in Eq.~(\ref{eq:eigen}). 
The equations in Eq. (\ref{eq:multi}) represent a cooling of the multi-basis states, and we simply call this new method ''multicool''.

In the multicool, one obtains the set of the AMD configurations $\{\Phi_n\}$ mainly for the ground state of a nucleus due to the energy minimization,
but does not optimize the configurations for the excited states. 
In the present study, we propose one method to obtain the configurations for the excited states in the multi AMD basis states.
It is naively expected that these configurations should be orthogonal to those for the ground state.
Considering this property, we employ the orthogonal projection method proposed by Kukulin et al. \cite{kukulin78}.
This method is often used in the orthogonality condition model (OCM) of the nuclear cluster systems
to remove the Pauli-forbidden states from the intercluster motions in the structures and reactions of nuclei \cite{aoyama01,myo14}.

We first fix the configurations $\{\Phi_n\}$ obtained in the multicool in Eq.~(\ref{eq:multi}) and regard them as $\{\Phi_{c_0}\}$ with the index $c_0=1,\cdots,N$.
We define the pseudo potential $V_\lambda$ using $\{\Phi_{c_0}\}$ in the projection form multiplying a real positive parameter $\lambda$
and add this potential to the Hamiltonian; 
\begin{equation}
  \begin{split}
    H_\lambda &= H + V_\lambda,\quad
    V_\lambda = \lambda \sum_{c_0=1}^N \kets{\Phi_{c_0}}\bras{\Phi_{c_0}},
    \\
    \Phi_\lambda &= \sum_{n=1}^N C_{\lambda,n} \Phi_{\lambda,n},\quad
    E_{{\rm GCM},\lambda} = \dfrac{ \bra \Phi_\lambda |H_\lambda| \Phi_\lambda \ket }{\bra \Phi_\lambda | \Phi_\lambda \ket}.
  \end{split}
  \label{eq:PSE}
\end{equation}
Using $\{\Phi_{c_0}\}$ with parity projection and a specific value of $\lambda$,
we perform the variation of the total energy $E_{{\rm GCM},\lambda}$.
We solve the multicool equation in Eq. (\ref{eq:multi}) for $H_\lambda$
and determine the basis states $\{\Phi_{\lambda,n}\}$ and the total wave function $\Phi_\lambda$. 
We increase the strength $\lambda$ starting from a small value and when $\lambda$ is sufficiently large,
$\Phi_\lambda$ can be orthogonal to $\{\Phi_{c_0}\}$. 
We take several values of $\lambda$ to trace the basis states $\{\Phi_{\lambda,n}\}$ including the transitional region from $\{\Phi_{c_0}\}$.
At each $\lambda$, we evaluate the total energy $E_{{\rm GCM},\lambda}$ subtracting the contribution of the pseudo potential,
although its contribution is minor.
The value of $\lambda$ can be as large as $10^5$-$10^6$ MeV in the typical OCM calculation of the multi-cluster systems \cite{aoyama01}.

In the multicool, the variation is performed for the intrinsic state without the angular momentum projection.
Due to this approximation, when the configuration $\Phi_{c_0}$ is spatially deformed to a certain direction,
the configuration $\Phi_{\lambda,n}$ can be deformed to another direction.
These configurations can be orthogonal to each other in the intrinsic frame,
but they might have a finite overlap after the angular momentum projection. 
To avoid this situation, we rotate the configurations $\{\Phi_{c_0}\}$ and add them to the projection term of the pseudo potential.
In this study, we adopt two kinds of rotations; $(x,y,z)\to(z,x,y)$ and $(x,y,z)\to(y,z,x)$ with parity projection
and the basis number becomes $3N$ in the pseudo potential.
We describe the configurations including rotations as $\{\Phi_{c}\}$ with the index $c=1,\cdots,3N$;
\begin{equation}
  \begin{split}
    V_\lambda &= \lambda \sum_{c=1}^{3N} \kets{\Phi_c}\bras{\Phi_c}.
  \end{split}
  \label{eq:PSE2}
\end{equation}
We use this potential to produce the multi-configurations for the excited states in the multicool.
The present orthogonal condition is a simplified case of the Schmidt's orthogonalization after the angular momentum projection \cite{kanada99,kanada98}.

We summarize the calculation procedure of the multicool for a nucleus with a mass number $A$.
\begin{enumerate}
\item
  We prepare the initial AMD basis states with a number $N$ setting the basis parameters randomly.
  The basis parameters are $\{X_{n,i}\}$ in Eq. (\ref{eq:multi}) for the basis index $n=1,\cdots,N$ and the particle index $i=1,\cdots,A$.
  The initial coefficients $\{C_n\}$ are determined by solving the eigenvalue problem of the Hamiltonian matrix in the intrinsic frame with parity projection.

\item
  Solving the multicool equation with the multi AMD basis states in Eq. (\ref{eq:multi}),
  we minimize the total energy and determine all the variational parameters $\{X_{n,i}\}$.
  The configurations $\{\Phi_n\}$ can be dominant for the ground state of a nucleus.
  We regard $\{\Phi_n\}$ as $\{\Phi_c\}$ with $c=1,\cdots,3N$ including the rotations of $\{\Phi_n\}$,
  used in the pseudo potential.
  
\item
  Putting a specific $\lambda$ in the pseudo potential in Eq.~(\ref{eq:PSE2}),
  we solve the multicool equation to obtain the configurations for the excited states.
  At each $\lambda$, we obtain the different set of configurations $\{\Phi_{\lambda,n}\}$.

\item
  Finally, we superpose $\{\Phi_n\}$ and $\{\Phi_{\lambda,n}\}$ for various $\lambda$ in GCM and
  solve the eigenvalue problem of the Hamiltonian matrix with the angular momentum projection
  in Eq. (\ref{eq:eigen}). We obtain the ground and excited states of a nucleus.
  It is noted that we treat $\lambda$ as a kind of the constraint parameter to generate the GCM basis states.
\end{enumerate}

\begin{figure*}[th]
\centering
\includegraphics[width=6.0cm,clip]{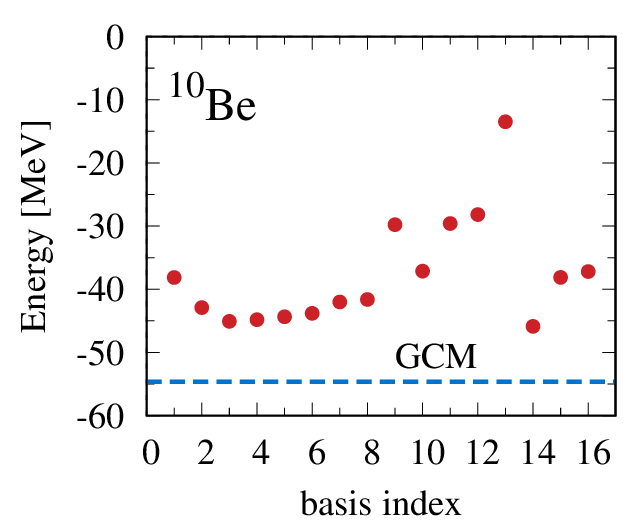}\hspace*{-0.05cm}
\includegraphics[width=6.0cm,clip]{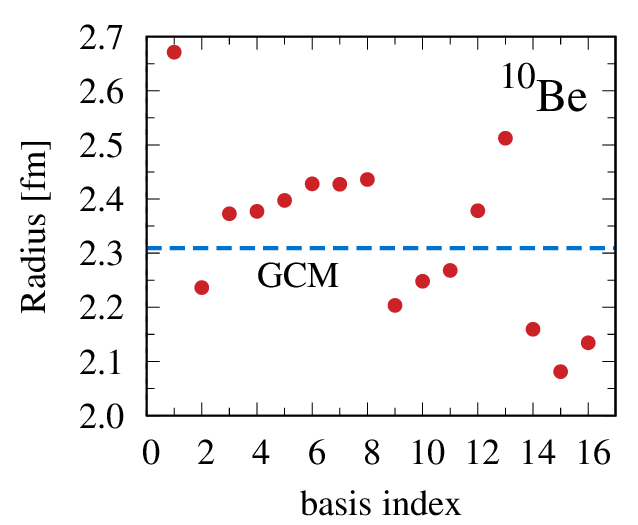}\hspace*{-0.05cm}
\includegraphics[width=6.0cm,clip]{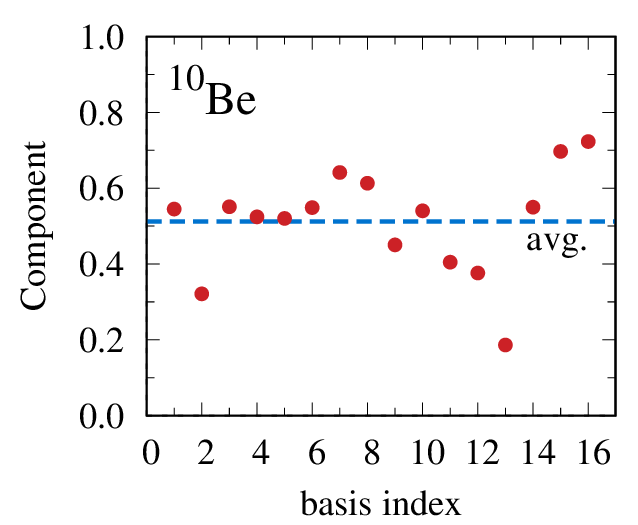}
\caption{
  Configurations of $^{10}$Be in the spin-fix case for the intrinsic positive parity state.
  The horizontal axis is the basis index $n$.
  Total energies in units of MeV (left), matter radii in units of fm (middle),
  and the components $| \langle \Phi_n| \Phi \rangle |^2$ (right) are shown with red dots for each configuration.
  The blue dashed lines represent the energy and radius of the total wave function with GCM and the average of the components.}
\label{fig:ene_fix}
\end{figure*}

\section{Results}\label{sec:result}

We perform the multicool calculation for $^{10}$Be with a positive parity state and discuss the reliability of the method.
We first prepare the initial AMD basis states for $^{10}$Be with $N=16$ and
this number is also used to produce the configurations for the excited states using the pseudo potential with various strengths of $\lambda$.
This number is sufficient to converge the final results of the ground and excited states of $^{10}$Be with GCM.

In the energy variation, we employ two schemes; one is that the directions of the intrinsic spin of nucleons
are fixed during the variation, called ''spin-fix'', and the other is that the spin directions are optimized during the variation, called ''spin-free''.
We finally superpose all the basis states obtained in the two schemes in the total wave function.
In the spin-fix case, we prepare the five nucleons with spin-up ($\alpha_i=1,\beta_i=0$) and spin-down ($\alpha_i=0,\beta_i=1$) directions, respectively, for $^{10}$Be.
In the two schemes, the obtained configurations are similar such as the cluster structures.
One difference is that the spin-free calculation tends to give the configurations of $^{10}$Be with smaller radii
than those of the spin-fix case. This is because considering two nucleons with the same spin direction,
they can come close to each other by the nucleon-nucleon interaction changing their spin directions to satisfy the Pauli principle.

\subsection{Spin-fix}\label{sec:spin_fix}

First, we discuss the spin-fix case.
We perform the multicool calculation to determine the ground-state configurations $\{\Phi_n\}$ for $^{10}$Be without the pseudo potential.
In Fig.~\ref{fig:ene_fix}, we show the total energies, matter radii, and components of each configuration of $^{10}$Be
for the intrinsic positive parity state,
in which the component of the basis index $n$ is defined as $|\langle \Phi_n|\Phi \rangle|^2$ in Eq.~(\ref{eq:multi}).
The total energy and radius of the GCM wave function $\Phi$ are shown in the blue dashed lines.
It is noted that the calculation results give the order of 16 configurations randomly,
and then we rearrange them in the order of the groups having similar structures such as density and clustering,
which are shown later, to make the discussion easier.

In the results, the energies of the configurations are distributed widely from $-45$ MeV to $-13$ MeV.
The GCM energy is $-55$ MeV with an energy gain of 10 MeV in the superposition of the basis states.
For the radii of the configurations, the values are also distributed widely from 2.1 fm to 2.7 fm and the GCM result is 2.3 fm.
For the components, all configurations contribute to the total wave function indicating the strong configuration mixing
and their average is 0.51 and the standard deviation is 0.13.

In Table \ref{tab:ene_fix}, we compare the results of the multicool with the single AMD basis state for $N$=1.
It is found that the effect of multi-configuration is 5.8 MeV in the total energy.
The radius becomes larger by 0.12 fm because of the mixture of the configurations having large radii.

\begin{figure*}[th]
\centering
\includegraphics[width=18.0cm,clip]{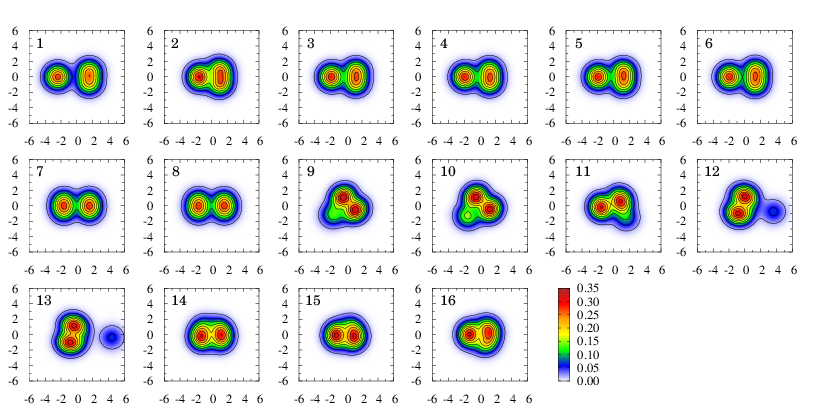}
\caption{
  Intrinsic density distributions of the configurations of $^{10}$Be in the spin-fix case.
  Units of densities and axes are fm$^{-3}$ and fm, respectively.
  The number in each panel means the basis index used in Fig. \ref{fig:ene_fix}.
}
\label{fig:density_fix}
\end{figure*}

In Fig.~\ref{fig:density_fix}, we show the intrinsic density distributions of the configurations of $^{10}$Be,
in which the integration of each distribution in three-dimensional space gives a mass number of $10$.
We assign the basis index of each configuration in the order of the groups having similar structures.
In the figure, we adjust the direction of the longest distribution to be the horizontal axis by calculating the principal axes.
Some of the distributions look similar to each other, but they have different directions of the longest distribution before adjustment of the direction.
Mixing of these basis states corresponds to the restoration of the rotational symmetry in GCM.
In the dense part with red color in the distributions, the $\alpha$ cluster often forms with two protons and two neutrons.

One can confirm various density distributions including the different clustering.
We classify them according to their spatial structures in Table \ref{tab:class_fix}.
The first six basis states with an index of 1--6 have the $^6$He+$\alpha$ configuration and
the 1st basis state gives the largest radius of 2.67 fm and the 2nd one gives the smallest radius of 2.24 fm among them.
This difference comes from the distance between $^6$He and $\alpha$, which indicates the GCM effect of the relative motion.
If the relative distance becomes small, the configuration approaches the shell-like state shown in the 16th basis state
with a small radius of 2.13 fm.

For the $^5$He+$^5$He configuration, the 7th and 8th basis states are assigned and if the relative distance between two $^5$He nuclei becomes small, 
the state approaches the symmetric shell-like state shown in the 14th basis state with a small radius of 2.16 fm.
This also indicates the GCM effect of the relative motion.
The $^8$Be+$2n$ configuration is also confirmed in two basis states of the 9th and 10th, and the dineutron is spatially enhanced in the 10th basis state.
The $^9$Be+$n$ configuration is confirmed in three basis states of 11th--13th with the different relative distances.
The last three basis states represent the shell-like configurations with small radii.

In the $^6$He+$\alpha$ and $^5$He+$^5$He configurations, the deformations of $^6$He and $^5$He are confirmed
in the vertical direction and perpendicular to the horizontal one with two clusters being lined.
In these configurations, according to the four-body picture of $\alpha$+$\alpha$+$n$+$n$,
we can interpret that two valence neutrons occupy the $p$-orbit around either of two $\alpha$s,
which corresponds to the $\pi$-orbit in the concept of the molecular orbital \cite{itagaki00,ito14}.

In the summary of the multicool calculation of $^{10}$Be in AMD, one confirms the various configurations with shell-like and clustering,
some of which show the different intercluster distances, representing the GCM effect of the relative motion.
This means that the preferable generator coordinates are determined automatically in the results of the energy variation of the total system.
This is the advantage of the multicool method to produce the appropriate multi-Slater determinants in AMD without any constraint.

\begin{table}[t]
\begin{center}
  \caption{
    Total energies and matter radii of the intrinsic ground state of $^{10}$Be with positive parity in the spin-fix case.
    Three calculations mean the single basis state, the multicool, and the multicool using the pseudo potential with the strength $\lambda$.
  }
\label{tab:ene_fix} 
\renewcommand{\arraystretch}{1.5}
\begin{tabular}{lccccccc}
\noalign{\hrule height 0.5pt}
                     &~~Energy [MeV]~~&~~Radius [fm]\\
\noalign{\hrule height 0.5pt}
Single                        &  $-48.9$     &  $2.20$  \\
Multicool                     &  $-54.7$     &  $2.31$  \\
Multicool,~$\lambda=10^5$ MeV &  $-48.0$     &  $2.90$  \\
\noalign{\hrule height 0.5pt}
\end{tabular}
\end{center}
\end{table}

\begin{table}[b]
\begin{center}
  \caption{
    Classification of the intrinsic density distributions of $^{10}$Be as shown in Fig. \ref{fig:density_fix} for the multiool
    and Fig. \ref{fig:density_fix_pse} for the multicool with $\lambda=10^5$ MeV in the pseudo potential. The number indicates the basis index.
    }
\label{tab:class_fix} 
\renewcommand{\arraystretch}{1.5}
\begin{tabular}{lllllll}
\noalign{\hrule height 0.5pt}
Configuration    && Multicool    &&&  Multicool, $\lambda=10^5$ MeV \\
\noalign{\hrule height 0.5pt}
$^6$He+$\alpha$  && 1,2,3,4,5,6  &&& 1,2,3,4,5,6    \\
$^5$He+$^5$He    && 7,8          &&& 7,8,9,10,11    \\
$^8$Be+$2n$      && 9,10         &&& 12             \\
$^9$Be+$n$       && 11,12,13     &&& 13,14,15       \\
shell-like       && 14,15,16     &&&                \\
$^7$Li+$^3$H     &&              &&& 16             \\
\noalign{\hrule height 0.5pt}
\end{tabular}
\end{center}
\end{table}

\begin{figure}[t]
\centering
\includegraphics[width=7.5cm,clip]{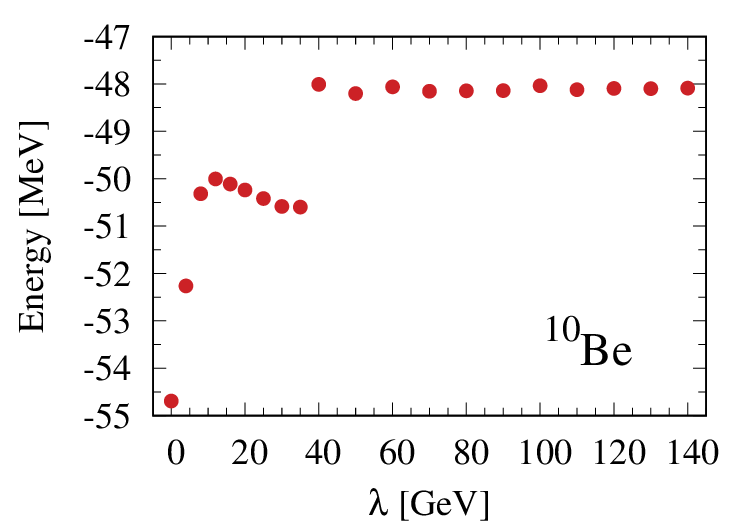}
\includegraphics[width=7.5cm,clip]{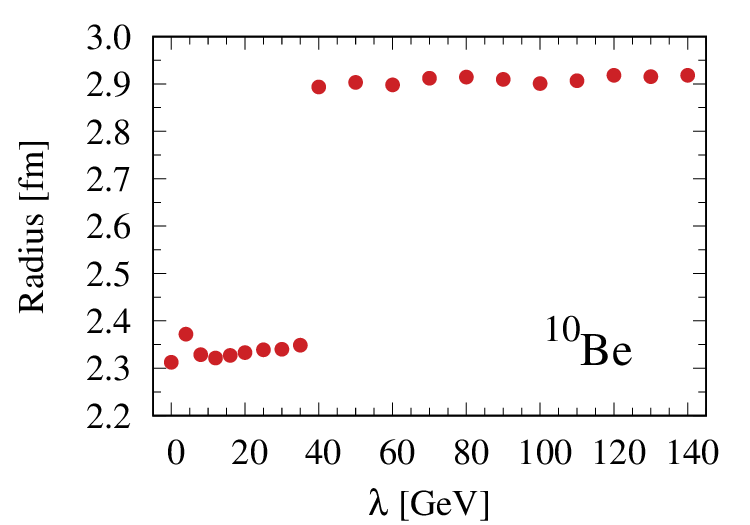}
\caption{
  Intrinsic energy and radius of the total wave function of $^{10}$Be with positive parity in the spin-fix case.
  The strength $\lambda$ of the pseudo potential changes.}
\label{fig:ene_fix_lam}
\end{figure}

We regard the AMD basis states $\{\Phi_n\}$ obtained in Eq. (\ref{eq:multi}) as the ground-state configurations for $^{10}$Be. 
In the pseudo potential in Eq.~(\ref{eq:PSE2}),
we adopt these configurations with rotations in the states $\{\Phi_c\}$
and produce the configurations for the excited states with a specific strength $\lambda$.
In Fig.~\ref{fig:ene_fix_lam}, we show the $\lambda$ dependences of the total energy and the radius of the total wave function of $^{10}$Be,
in which the contribution of the pseudo potential is removed from the total energy.
At each $\lambda$, the expectation value of the pseudo potential,
$\langle V_\lambda \rangle=\lambda \sum_{c=1}^{3N} |\langle \Phi_c|\Phi_\lambda \rangle|^2 $ is typically less than 0.5 MeV
including the large prefactor $\lambda$,
which means that $\sum_{c=1}^{3N} |\langle \Phi_c|\Phi_\lambda \rangle|^2 $ is tiny.
This is sufficient to obtain the configurations $\{\Phi_{\lambda,n}\}$ different from $\{\Phi_c\}$.

By increasing $\lambda$ from zero, the total energy goes up first until $\lambda=10^4$ MeV by about 5 MeV,
and after that slightly goes down and again goes up rapidly at $\lambda=4\times 10^4$ MeV
and becomes the stable energy of $-48$ MeV with the excitation energy of around 7 MeV.
For the radius, the same tendency is confirmed and the excited states show a stable radius of 2.9 fm,
which is extended from 2.3 fm of the ground state.
These results indicate the structural transition of $^{10}$Be after $\lambda=4\times 10^4$ MeV.

\begin{figure*}[th]
\centering
\includegraphics[width=6.0cm,clip]{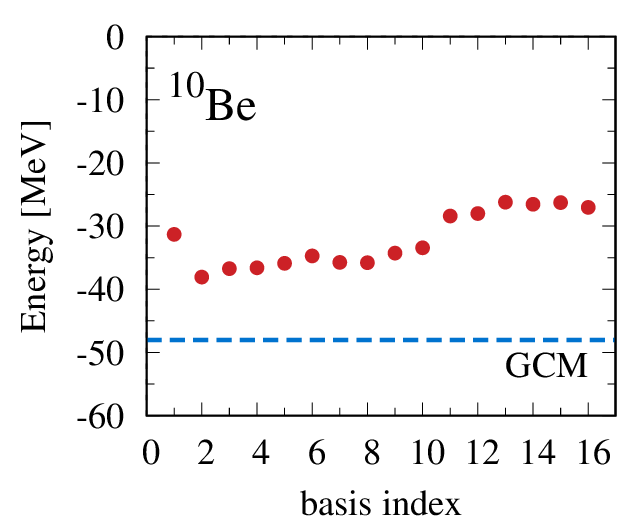}\hspace*{-0.05cm}
\includegraphics[width=6.0cm,clip]{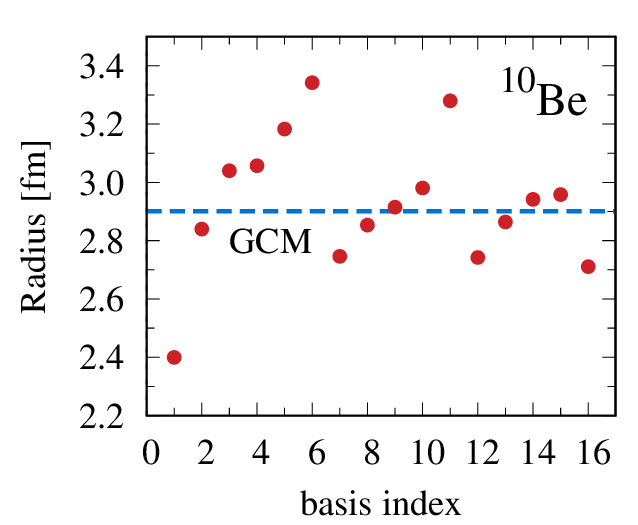}\hspace*{-0.05cm}
\includegraphics[width=6.0cm,clip]{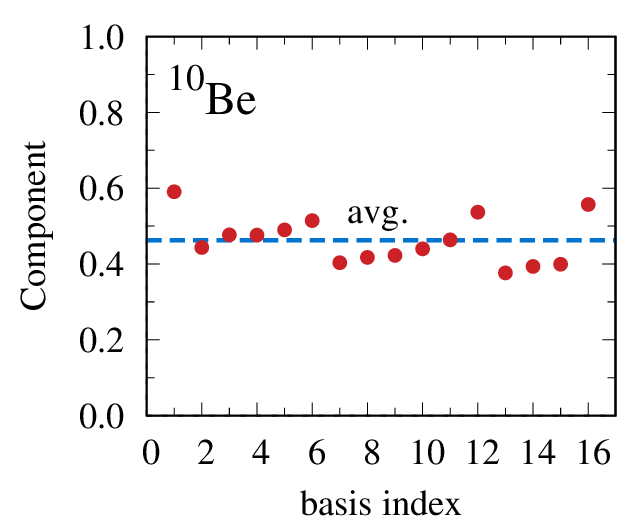}
\caption{
  Configurations of $^{10}$Be in the spin-fix case for the intrinsic positive parity state.
  The strength $\lambda=10^5$ MeV is used in the pseudo potential. The horizontal axis represents the basis index $n$.
  Total energies in units of MeV (left), matter radii in units of fm (middle), and the components $| \langle \Phi_{\lambda,n}| \Phi_\lambda \rangle |^2$ (right)
  are shown with red dots for each configuration.
  The blue dashed lines represent the energy and radius of the total wave function with GCM and the average of the components.}
\label{fig:ene_fix_PSE}
\end{figure*}

For the excited states of $^{10}$Be, we discuss the configurations with $\lambda=10^5$ MeV.
We show the total energies, radii, and the components in the intrinsic frame in Fig. \ref{fig:ene_fix_PSE}.
The energies are distributed from $-38$ MeV to $-26$ MeV, and the radii are also widely distributed from 2.4 fm to 3.4 fm.
For the components, all configurations contribute to the total wave function with similar values of 0.4--0.6,
indicating the strong configuration mixing and their average is 0.46 with a small standard deviation of 0.06.
These tendencies are common for the case as shown in Fig. \ref{fig:ene_fix},
and the coherency of the configurations is stronger in the excited state than that in the ground state from the components.

\begin{figure*}[th]
\centering
\includegraphics[width=18.0cm,clip]{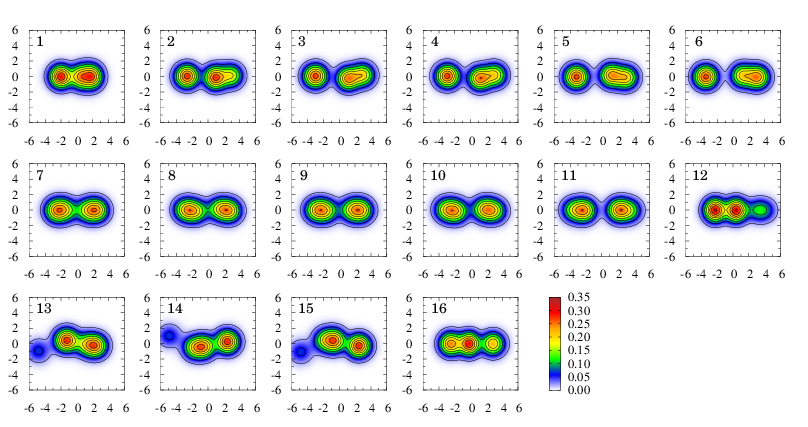}
\caption{
  Intrinsic density distribution of each configuration of $^{10}$Be
  in the spin-fix case using the pseudo potential with the strength of $\lambda=10^5$ MeV.
  Units of densities and axes are fm$^{-3}$ and fm, respectively.
  The number in each panel means the basis index used in Fig. \ref{fig:ene_fix_PSE}.
}
\label{fig:density_fix_pse}
\end{figure*}

In Fig. \ref{fig:density_fix_pse}, we show the density distributions of the excited-state configurations of $^{10}$Be,
which are significantly different from those for the ground-state configurations shown in Fig.~\ref{fig:density_fix}.
It is found that all configurations commonly have an elongate structure with various clustering like a linear chain.
In Table \ref{tab:class_fix}, we classify the structure of each configuration similarly to the calculation for the ground-state configurations.
The first six states show the $^6$He+$\alpha$ configuration and their structures show the variety;
in the 1st basis state, the small relative distance between $^6$He and $\alpha$ is confirmed with the smallest radius of 2.4 fm among all the basis states.
Two valence neutrons are located in between two $\alpha$s and are close to the right-hand side of $\alpha$ to form $^6$He.
In the 2nd and 3rd basis states, $^6$He and $\alpha$ are spatially separated and two valence neutrons in $^6$He are located at the right end of the density.
In the 6th basis state, $^6$He and $\alpha$ are the most separated and the radius shows the largest value of 3.35 fm among all the basis states.
This basis state corresponds to the extension of the relative motion from the 1st basis state.

\begin{figure}[th]
\centering
\includegraphics[width=6.5cm,clip]{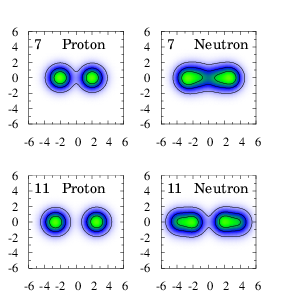}
\caption{
  Intrinsic density distributions of protons and neutrons in the 7th (upper) and 11th (lower) configurations of $^{10}$Be
  with the same condition given in Fig. \ref{fig:density_fix_pse}.
}
\label{fig:density_fix_pse_7_11}
\end{figure}

Five basis states from 7th to 11th show the $^5$He+$^5$He configuration with various relative distances, which represents the GCM effect.
Among them, the 7th basis state gives the smallest radius of 2.75 fm and the 11th basis state gives the largest radius of 3.3 fm.
It is noted that the direction of deformation of two $^5$He nuclei is commonly along the horizontal axis
and is different from that in the ground-state configurations shown in Fig. \ref{fig:density_fix}.
The location of two valence neutrons also depends on the basis states; mainly in between two $\alpha$s (7th) and on the left and right ends of the density (11th) along the horizontal axis.
To confirm it clearly, in Fig. \ref{fig:density_fix_pse_7_11}, the proton and neutron density distributions of the two basis states are shown
and one can realize the difference in the neutron distributions.
In the proton part, the different distances between the proton pairs are confirmed.

In the 12th basis state, two neutrons are spatially separated from two $\alpha$s contacting each other as $^8$Be, indicating the $^8$Be+$2n$ structure.
The $^9$Be+$n$ configuration is also confirmed in the 13th, 14th, and 15th basis states,
in which the last neutron is located at the left end of the density with a slight bending.
In the distributions, $^9$Be shows the $^5$He+$\alpha$ structure and the valence neutron is located in between two $\alpha$s (13th) and on the left end of $^9$Be (14th, 15th).
The last 16th basis state shows the $^7$Li+$^3$H (triton) configuration where $^7$Li shows the $\alpha$+$^3$H structure contacting with each other. 
This means that one $\alpha$ cluster and two tritons form in $^{10}$Be.
We do not confirm the spatially compact shell-like state among the excited-state configurations.

According to the four-body picture of $\alpha$+$\alpha$+$n$+$n$, in the configurations of $^6$He+$\alpha$, $^5$He+$^5$He, and $^8$Be+$2n$,
two valence neutrons mostly occupy the orbit in the horizontal direction along with two $\alpha$s,
with various locations and various relative distances between two $\alpha$s.
We can interpret this structure of two neutrons as the $\sigma$-orbit in the molecular orbital picture \cite{itagaki00}.
From the results of the excited-state configurations together with the ground-state ones,
the molecular orbital structure is automatically confirmed in $^{10}$Be in the results of variation of multi-Slater determinants without a priori.

\subsection{Spin-free}\label{sec:spin_free}

We perform the multicool calculation of $^{10}$Be with the variation of the spin directions of all nucleons in the AMD basis states.
The basis number is 16 and is common as used in the spin-fix case.
In Table \ref{tab:ene_free}, we obtain the ground-state energy in the multicool and compare the results with the single AMD basis state.
The single basis state gives a total energy of $-49.5$ MeV and the multicool gives $-56.0$ MeV, and the energy gain is 5.5 MeV.
In the multicool, the radius is 2.23 fm, which is smaller than the spin-fix case of 2.31 fm, as shown in Table \ref{tab:ene_fix}.
This is because two nucleons can come close to each other by the interaction changing their spin directions.
This effect gains the total energy of $^{10}$Be by $1.3$ MeV in comparison with the spin-fix case of $-54.7$ MeV.

\begin{figure}[th]
\centering
\includegraphics[width=7.5cm,clip]{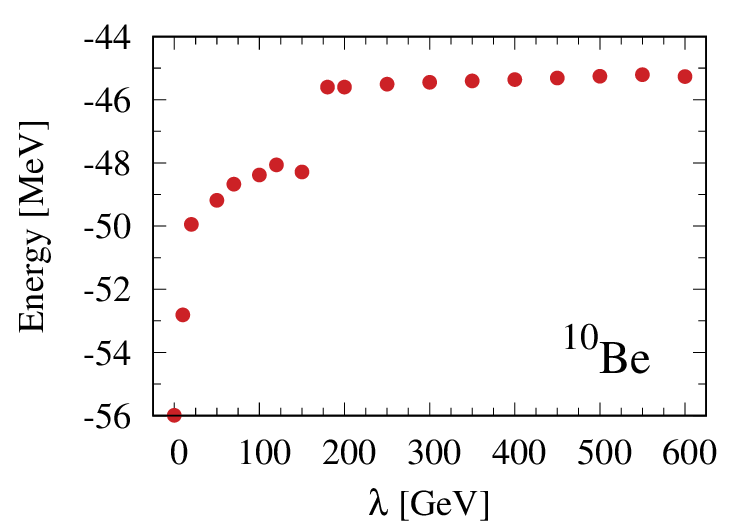}
\includegraphics[width=7.5cm,clip]{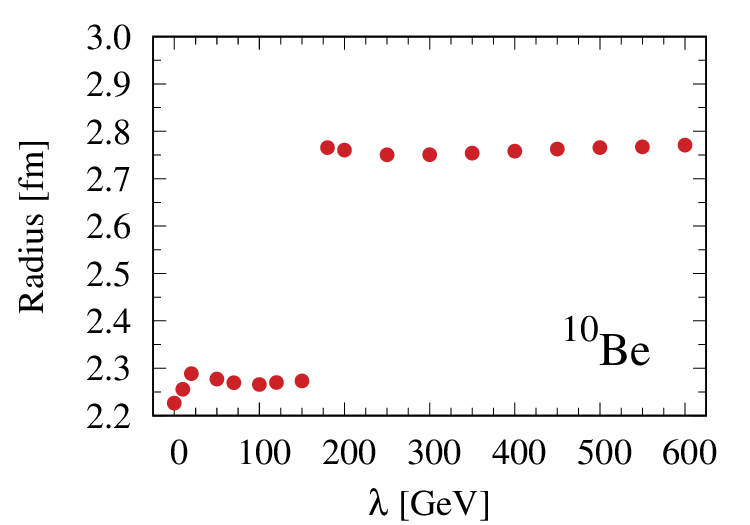}
\caption{
  Intrinsic energy and radius of the total wave function of $^{10}$Be with positive parity in the spin-free case.
  The strength $\lambda$ of the pseudo potential changes.}
\label{fig:ene_free_lam}
\end{figure}

We add the pseudo potential in the Hamiltonian with the strength $\lambda$ using the ground-state configurations in the multicool
and produce the configurations for the excited states.
In Fig.~\ref{fig:ene_free_lam}, we show the results of changing the strength $\lambda$ for the total energy and the radius in the multicool.
The overall trend is similar to the spin-fix case as shown in Fig.~\ref{fig:ene_fix_lam}.
When we use the value of $\lambda$ larger than $1.8\times 10^5$ MeV, the state transits to the stable solution with the higher energy and the larger radius.
As shown in Table \ref{tab:ene_free}, the excitation energy is about 11 MeV from $-56$ MeV to $-45$ MeV,
and the radius becomes as large as 2.75 fm, which is a little smaller than 2.9 fm of the spin-fix case in Fig.~\ref{fig:ene_fix_lam}.

In the spin-free calculation of $^{10}$Be, the densities of the ground and excited-state configurations 
provide similar distributions to those of the spin-fix case as shown in Figs. \ref{fig:density_fix} and \ref{fig:density_fix_pse},
such as the shell-like and clustering structures. 

In the spin-free case, the lowest value of $\lambda$ to get the excited state of $^{10}$Be 	with a large radius is larger than that of the spin-fix case.
This is because the variation of the spin directions of nucleons in $\Phi_\lambda$ can reduce
the overlap of $\langle \Phi_c|\Phi_\lambda \rangle$ in the matrix elements of the pseudo potential without the spatial extension of $\Phi_\lambda$.
Hence a larger $\lambda$ is necessary to extend the system spatially in the spin-free calculation.

\begin{table}[th]
\begin{center}
  \caption{
    Total energies and matter radii of the intrinsic ground state of $^{10}$Be with positive parity in the spin-free case.
    Three calculations mean the single basis state, the multicool, and the multicool using the pseudo potential with the strength $\lambda$.
    }
\label{tab:ene_free} 
\renewcommand{\arraystretch}{1.5}
\begin{tabular}{lccccccc}
\noalign{\hrule height 0.5pt}
                                      &~~Energy [MeV]~~&~~Radius [fm]\\
\noalign{\hrule height 0.5pt}
Single                                &  $-49.5$     &  $2.18$  \\
Multicool                             &  $-56.0$     &  $2.23$  \\
Multicool,~$\lambda=3\times 10^5$ MeV &  $-45.5$     &  $2.75$  \\
\noalign{\hrule height 0.5pt}
\end{tabular}
\end{center}
\end{table}

\subsection{Energy spectrum}\label{sec:level}

We perform the angular momentum projection of the multicool basis states of $^{10}$Be for the positive parity states
and solve the eigenvalue problem of the Hamiltonian matrix in Eq.~(\ref{eq:eigen}).
This calculation corresponds to the GCM one using the basis states obtained in the multicool.
For spin-fix and spin-free cases, we employ all the basis states obtained for the ground state and the excited states with the different values of $\lambda$,
as shown in Figs. \ref{fig:ene_fix_lam} and \ref{fig:ene_free_lam}.
The total basis number is about 600 in the present calculation.

Here, we mention the threshold energies to separate $^{10}$Be to the subsystems.
We calculate $^6$He and $^9$Be in the multicool method in a similar way to $^{10}$Be.
The energy of the $0^+$ ground state of $^6$He is obtained as $-29.2$ MeV with the two-neutron separation energy is 1.6 MeV
measured from the energy of $-27.6$ MeV of the $\alpha$ particle, and the $^6$He+$\alpha$ threshold energy is $-56.8$ MeV.
The energy of the $3/2^-$ ground state of $^9$Be is obtained as $-57.5$ MeV, which becomes the threshold energy of $^9$Be+$n$.
We summarize these values in Table \ref{tab:ene_sub} with their radii in comparison with the experimental values. 
The charge radii of $^6$He and $^9$Be are consistent with the experimental values \cite{mueller07,nortershauser09}.

\begin{table}[t]
\begin{center}
  \caption{
    Total energies and radii of $^6$He and $^9$Be.
    Radii of matter, proton, neutron, and charge are $r_{\rm m}$, $r_{\rm p}$, $r_{\rm n}$, and $r_{\rm ch}$, respectively.
    The values with square brackets are the experimental energies and charge radii \cite{mueller07,nortershauser09}.
    Units of energy and radius are MeV and fm, respectively.}
\label{tab:ene_sub}
\renewcommand{\arraystretch}{1.5}
\begin{tabular}{llllllllllll}
\noalign{\hrule height 0.5pt}
                &~~Energy~~          &&~$r_{\rm m}$  &&~$r_{\rm p}$ &&~$r_{\rm n}$ &&~$r_{\rm ch}$ \\
\noalign{\hrule height 0.5pt}
$^6$He($0^+$)   &  $-29.2$ [$-29.3$]  &&  2.38 && 1.88  && 2.59 && 2.04 [2.068(11)] \\
$^9$Be($3/2^-$) &  $-57.5$ [$-58.2$]  &&  2.44 && 2.37  && 2.50 && 2.51 [2.519(12)] \\
\noalign{\hrule height 0.5pt}
\end{tabular}
\end{center}
\end{table}

\begin{figure*}[t]
\centering
\includegraphics[width=14.0cm,clip]{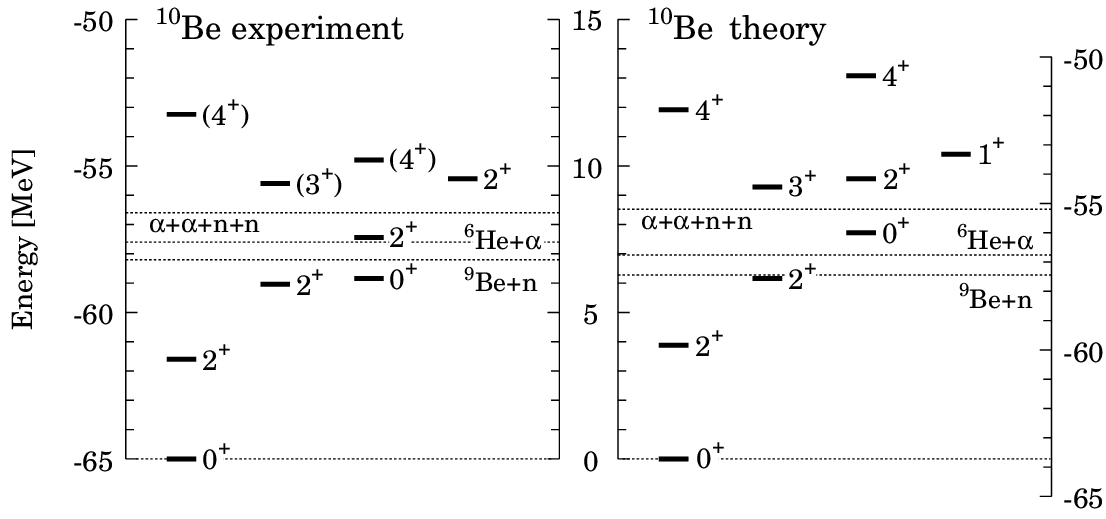}
\caption{
  Energy spectrum of $^{10}$Be for the experiment (left) \cite{tilley04} and the multicool (right) in units of MeV.
  The left-hand and right-hand sides of the measures represent the total energies and the measures in the middle represent the excitation energy.
  The threshold energies of $\alpha$+$\alpha$+$n$+$n$, $^6$He+$\alpha$, and $^9$Be+$n$ are shown with dotted horizontal lines.}
\label{fig:ene_GCM}
\end{figure*}

For $^{10}$Be, the energy spectrum is shown in Fig. \ref{fig:ene_GCM} in comparison with the experiments.
The present results fairly agree with the observed spectrum for positive parity states.
In Table \ref{tab:ene_GCM}, the total energies and radii of each state are listed.
For the $0^+$ states, the $0^+_1$ state is spatially compact with a matter radius of 2.33 fm
and is mainly described by the configurations with small values of $\lambda$ before the structural transition of $^{10}$Be. 
The total energy is $-63.7$ MeV, which is close to the experimental value of $-64.98$ MeV. 
For the spin-fix basis states only, the energy is $-62.5$ MeV and the spin-free case is $-62.8$ MeV,
and their coupling gains the energy by about 1 MeV.

The $0^+_2$ state is spatially extended with a radius of 2.88 fm and is mainly described by the excited-state configurations
with large radii produced using large values of $\lambda$ after the transition.
The $0^+_2$ state is obtained lower than the $\alpha$+$\alpha$+$n$+$n$ threshold energy of $-55.2$ MeV by about 0.8 MeV.
The excitation energy is $7.7$ MeV, which is close to, but slightly higher than the experimental value of $6.2$ MeV by 1.5 MeV.
This state is mainly described by the basis states in the spin-fix case with large radii.
In the spin-fix basis states only, the energy of the $0^+_2$ state is $-55.7$ MeV, which is close to $-56.0$ MeV in the full basis states,
with a radius of 2.89 fm.
The spin-free basis states only give the energy of $-52.1$ MeV with a radius of 2.82 fm.

We predict the $1^+$ state with the excitation energy of 10 MeV, which is not observed experimentally, and predicted in other theories \cite{itagaki00,carlson15}.
For $2^+$ states, the $2^+_1$ and $2^+_2$ states are rather spatially compact states.
The $2^+_3$ state gives a large radius and mainly described by the excited-state configurations with large radii.
This state can be the band member of the $0^+_2$ state because the two states have a similar large radius of 2.9 fm.

We obtain one $3^+$ state, which can be the member of $K^\pi=2^+$ band with $2^+_2$ showing a similar radius of around 2.4 fm \cite{itagaki00,suhara10}.
The candidate of this state is reported experimentally with a similar excitation energy of 9.4 MeV \cite{tilley04}.  
We obtain two $4^+$ states and the $4^+_1$ state can be the band member of the $0^+_1$ and $2^+_1$
and three states give a similar radius of around 2.3--2.4 fm.
The $4^+_2$ state shows a large radius and can be the band member of the $0^+_2$ and $2^+_3$ and
three states also give a similar radius of around 2.9 fm.

\begin{table}[t]
\begin{center}
  \caption{
    Total energies of the $^{10}$Be states in the multicool in units of MeV.
    Matter radii are shown in units of fm.
  }
\label{tab:ene_GCM}
\renewcommand{\arraystretch}{1.5}
\begin{tabular}{clllllllllll}
\noalign{\hrule height 0.5pt}
        &&&  ~~Energy [MeV] &&& Radius [fm]  \\
\noalign{\hrule height 0.5pt}
$0^+_1$ &&&  $-63.72$ &&& $2.33$ \\
$0^+_2$ &&&  $-56.01$ &&& $2.88$ \\
\noalign{\hrule height 0.5pt}       
$1^+$   &&&  $-53.31$ &&& $2.51$ \\ 
\noalign{\hrule height 0.5pt}       
$2^+_1$ &&&  $-59.84$ &&& $2.33$ \\
$2^+_2$ &&&  $-57.54$ &&& $2.42$ \\
$2^+_3$ &&&  $-54.16$ &&& $2.89$ \\
\noalign{\hrule height 0.5pt}       
$3^+ $  &&&  $-54.25$ &&& $2.40$ \\
\noalign{\hrule height 0.5pt}       
$4^+_1$ &&&  $-51.99$ &&& $2.40$ \\
$4^+_2$ &&&  $-50.64$ &&& $2.91$ \\
\noalign{\hrule height 0.5pt}
\end{tabular}
\end{center}
\end{table}

We discuss the GCM effect on the energies of $^{10}$Be in the multicool and focus on the ground $0^+$ state.
In the single AMD basis state, the lowest intrinsic energy is $-49.5$ MeV as shown in Table \ref{tab:ene_free},
and after the angular momentum projection, the $0^+$ state gives $-54.9$ MeV with the energy gain of $5.4$ MeV.
In the multicool with 16 configurations, the lowest intrinsic energy is $-56.0$ MeV in the spin-free case.
The superposition of these basis states with the angular momentum projection gives the energy of $-60.7$ MeV 
and by adding more 16 configurations in the spin-free case, we obtain the energy of $-61.8$ MeV.
Hence the projection effect is $5.8$ MeV in the multicool, which is close to the value obtained in the single basis state.
Using the pseudo potential, we construct the various configurations and finally adding these configrations we obtain the energy of $-63.7$ MeV.
This means that the effect of the configuration mixing with the pseudo potential is $1.9$ MeV,
which is smaller than the effect of the angular momentum projection.

For reference, in the $2^+_1$ state, the single basis state gives the $-51.8$ MeV, and the multicool without the pseudo potential gives the $-58.0$ MeV,
and the final energy with all configurations is $-59.8$ MeV.

In Table \ref{tab:ene_cmp}, we compare the present energies of $^{10}$Be with the four kinds of similar cluster models;
Molecular orbital model of $\alpha$+$\alpha$+$n$+$n$ (MO) assuming two $\alpha$ clusters \cite{itagaki00},
AMD with the constraints of the $\beta$--$\gamma$ deformations \cite{suhara10},
AMD with dineutron condensation model (DC) \cite{kobayashi11}, and
AMD with the $\beta$--$\gamma$ constraints in the $K$ quantum number projection ($\beta$-$\gamma K$) \cite{shikata20}.
In these models and the present multicool, the same Hamiltonian is used except for MO with 2000 MeV of the LS strength \cite{itagaki00}.

It is found that the present multicool provides the lowest energy of the $0^+_1$ state with $-63.7$ MeV,
which is slightly lower than $-63$ MeV of $\beta$-$\gamma K$ \cite{shikata20}.
This result is consistent from the viewpoint of the variational principle.
It is found that $0^+_2$ state with the energy of $-56.0$ MeV
is lower than the values of other theories by about 5 MeV with the same Hamiltonian including the LS strength,
and also gives the lowest excitation energy of 7.7 MeV.
This indicates that the present multicool with the pseudo potential produces the appropriate configurations for the excited states.
When we use 2000 MeV of the LS strength, the energy of the $0^+_1$ ($0^+_2$) state is $-66.8$ MeV ($-57.3$ MeV)
and the excitation energy of the $0^+_2$ state is 9.5 MeV, which becomes larger than the present 7.7 MeV.
This is because of the larger attraction of the LS force in the $0^+_1$ state than that in the $0^+_2$ state,
which indicates the $jj$-coupling component in the $0^+_1$ state.
For the $2^+$, $3^+$, and $4^+$ states, our results are almost consistent with those of Refs. \cite{itagaki00,suhara10},
although the total energy is lower in the multicool entirely by about 4--5 MeV.

\begin{table}[t]
\begin{center}
  \caption{
    Excitation energies of $^{10}$Be in the multicool in comparison with four theories using the same Hamiltonian
    \cite{itagaki00,suhara10,shikata20,kobayashi11} except for \cite{itagaki00} with 2000 MeV of the LS strength.
    Units are MeV. The values with parentheses mean the total energies.
    }
  \label{tab:ene_cmp}
\renewcommand{\arraystretch}{1.5}
\begin{tabular}{clllllllllllll}
\noalign{\hrule height 0.5pt}
        &&  Multicool    && MO\cite{itagaki00}  && $\beta$-$\gamma$\cite{suhara10} && DC\cite{kobayashi11} && $\beta$-$\gamma K$\cite{shikata20} \\
\noalign{\hrule height 0.5pt}
$0^+_1$ &&  $0  $        && $0$         &&  $0        $ && $0$           &&  $0     $  \vspace*{-0.2cm}\\
        &&  $(-63.7)   $ && $(-61.4)$   &&  $(-59.2)  $ && $(-60.4)$     &&  $(-63) $  \\
$0^+_2$ &&  $7.7$        && $8.1$       &&  $8.0$       && $9.5$         &&  $12$  \vspace*{-0.2cm}\\
        &&  $(-56.0)$    && $(-53.3)$   &&  $(-51.2)$   && $(-50.9)$     &&  $(-51)$  \\
\noalign{\hrule height 0.5pt}
$1^+$   &&  $10.4$       && $10.1$       &&  --         && -- && -- \\
\noalign{\hrule height 0.5pt}
$2^+_1$ &&  $3.9$        && $3.3$         &&  $3.3$  && -- && -- \\
$2^+_2$ &&  $6.2$        && $5.7$         &&  $5.8$  && -- && -- \\
$2^+_3$ &&  $9.6$        && $9.5$         &&  $9.9$  && -- && -- \\
\noalign{\hrule height 0.5pt}
$3^+ $  &&  $9.5$        && $9.6$         &&  $9.2$  && -- && -- \\
\noalign{\hrule height 0.5pt}
$4^+_1$ &&  $11.7$       && $10.6$        &&  $11.0$ && -- && -- \\
$4^+_2$ &&  $13.1$       && $12.5$        &&  $13.5$ && -- && -- \\
\noalign{\hrule height 0.5pt}
\end{tabular}
\end{center}
\end{table}

In Table \ref{tab:radius}, we list the radial properties of two $0^+$ states of $^{10}$Be in comparison with the experimental and theoretical values.
For the $0^+_1$ state, the multicool gives 2.33 fm of matter radius, which is close to the experimental value \cite{ozawa01} and is smaller than other theories.
The charge radius in the multicool is consistent with the experimental value \cite{nortershauser09}.
For the $0^+_2$ state, our results give the large radius, in which both proton and neutron radii enhance from those of the $0^+_1$ state by about 0.5 fm.
This is the common feature seen in other theories.

\begin{table}[t]
\begin{center}
  \caption{
    Radii of the $0^+_{1,2}$ states of $^{10}$Be for matter ($r_{\rm m}$), proton ($r_{\rm p}$), neutron ($r_{\rm n}$), and charge ($r_{\rm ch}$)
    in comparison with the experiments \cite{nortershauser09,ozawa01}
    and other theories using the same Hamiltonian.
    Units are fm.}
\label{tab:radius}
\renewcommand{\arraystretch}{1.50}
\begin{tabular}{cllllllllllll}
\noalign{\hrule height 0.5pt}
              &&         && Expt.       && Multicool && MO\cite{itagaki00} && $\beta$-$\gamma$\cite{suhara10} && DC\cite{kobayashi11} \\
\noalign{\hrule height 0.5pt}
              && $r_{\rm m}$  &&  2.30(2)    &&  2.33    &&  --    &&  2.39     && 2.37 \\
 \lw{$0^+_1$} && $r_{\rm p}$  &&   --        &&  2.21    &&  --    &&  --       && 2.31 \\
              && $r_{\rm n}$  &&   --        &&  2.40    &&  --    &&  --       && 2.41 \\
              && $r_{\rm ch}$ &&  2.357(18)  &&  2.36    &&  2.51  &&  --       && -- \\ 
\noalign{\hrule height 0.5pt}                                  
              && $r_{\rm m}$  &&   --        &&  2.88    &&  --    &&  2.98     && 2.96 \\
 \lw{$0^+_2$} && $r_{\rm p}$  &&   --        &&  2.70    &&  --    &&  --       && 2.66 \\
              && $r_{\rm n}$  &&   --        &&  2.99    &&  --    &&  --       && 3.14 \\
              && $r_{\rm ch}$ &&   --        &&  2.82    &&  2.93  &&  --       && -- \\ 
\noalign{\hrule height 0.5pt}
\end{tabular}
\end{center}
\end{table}

Finally, in Table \ref{tab:E2}, we list the electric quadrupole transition strength $B(E2)$ from the $2^+$ states to the $0^+$ states,
some of which are compared with the experimental and theoretical values.
In the multicool, we almost obtained consistent results with other theories.
For $2^+_3$ to $0^+_2$, these states commonly have large radii, and the $E2$ strength shows a large value,
which is consistent with the results in MO \cite{itagaki00}.

\begin{table}[t]
\begin{center}
  \caption{
    Electric quadrupole transition strength $B(E2)$ of $^{10}$Be in the multicool
    in comparison with those of the experiments \cite{tilley04,mccutchan09}
    and other theories with the same Hamiltonian. Units are $e^2 {\rm fm}^4$.}
\label{tab:E2}
\renewcommand{\arraystretch}{1.5}
\begin{tabular}{cllllllllllll}
\noalign{\hrule height 0.5pt}
                 &&& Experiment            &&&  Multicool  &&& MO\cite{itagaki00}  &&& $\beta$-$\gamma$\cite{suhara10} \\
\noalign{\hrule height 0.5pt}
$2^+_1\to 0^+_1$ &&&  $10.5(1.0)$,~$9.2(3)$&&&  7.9    &&&  11.26 &&&  9.4  \\
$2^+_1\to 0^+_2$ &&&  $0.66(24)$           &&&  0.09   &&&   0.23 &&&  1.2  \\
\noalign{\hrule height 0.5pt}                                   
$2^+_2\to 0^+_1$ &&&  $0.11(2)$            &&&  0.37   &&&   0.44 &&&  0.7  \\
$2^+_2\to 0^+_2$ &&&   --                  &&&$<$0.002 &&&   0.00 &&&  --   \\
\noalign{\hrule height 0.5pt}                                   
$2^+_3\to 0^+_1$ &&&   --                  &&&  0.05   &&&   0.19 &&&  --   \\
$2^+_3\to 0^+_2$ &&&   --                  &&&  27.1   &&&  35.56 &&&  --   \\
\noalign{\hrule height 0.5pt}
\end{tabular}
\end{center}
\end{table}
\section{Summary}\label{sec:summary}

We proposed a method to optimize the multi-Slater determinants of the antisymmetrized molecular dynamics (AMD) in the linear combination form for nuclei.
The configurations of the Slater determinants and their weights are determined simultaneously to minimize the energy of the total wave function.
This method is useful to find the important generator coordinates for many-body systems.
We further optimize the excited-state configurations imposing the orthogonal condition to the ground-state configurations.
In this paper, we formulated this method in AMD and applied it to neutron-rich $^{10}$Be,
which shows the various structures of the shell-like and clustering states. 

In AMD, the nucleon wave functions are Gaussian wave packets and
the centroid parameters of Gaussians in all the Slater determinants are determined in the variation for the total wave function.
We employ the cooling method in the multi AMD basis states, which we call the multicool.
We first obtain the ground-state configurations 
and next, obtain the excited-state configurations under the orthogonal condition to the ground-state configurations.

In the multicool calculation of $^{10}$Be,
we obtain the basis states with the various spatial configurations and deformations including clustering of $^5$He+$^5$He, $^6$He+$\alpha$, $^7$Li+$^3$H, $^8$Be+$2n$, and $^9$Be+$n$.
These configurations are produced without a priori and support the molecular orbital picture of two valence neutrons surrounding two $\alpha$s in $^{10}$Be.
We also confirm that the appropriate generator coordinates, such as the intercluster distances,
are obtained automatically in the energy variation of the total system. This is the advantage of the present multicool.
By superposing all the basis states, we finally obtain the energy spectrum of $^{10}$Be, which reproduces the experiments.
We also provide the electric quadrupole transitions, which are consistent with those of other cluster models.
These results indicate the reliability of the present multicool.

The concept of the multicool is general for nuclei and it is interesting to apply it to the configuration mixing phenomena including the clustering.
We plan to extend this method to the ab initio type calculations, such as TOAMD and HM-AMD with the realistic nuclear forces,
to determine the optimal variational parameters in many basis states efficiently.
It is a remaining issue to describe the scattering states with resonances above the particle threshold energy.
One possibility for it is to combine the multicool with the complex scaling \cite{myo14,myo23}.

\section*{Acknowledgments}
We would like to thank Prof. Hiroshi Toki and Prof. Kiyoshi Kat\=o for their useful discussions and comments.
This work was supported by JSPS KAKENHI Grants No. JP22K03643.
One of us (M.L.) acknowledges the supports from the National Natural Science Foundation of China (Grants No. 12105141),
and the Jiangsu Provincial Natural Science Foundation (Grants No. BK20210277).

\section*{References}
\def\JL#1#2#3#4{ {{\rm #1}} \textbf{#2}, #4 (#3)}  
\nc{\PR}[3]     {\JL{Phys. Rev.}{#1}{#2}{#3}}
\nc{\PRC}[3]    {\JL{Phys. Rev.~C}{#1}{#2}{#3}}
\nc{\PRA}[3]    {\JL{Phys. Rev.~A}{#1}{#2}{#3}}
\nc{\PRL}[3]    {\JL{Phys. Rev. Lett.}{#1}{#2}{#3}}
\nc{\NP}[3]     {\JL{Nucl. Phys.}{#1}{#2}{#3}}
\nc{\NPA}[3]    {\JL{Nucl. Phys.}{A#1}{#2}{#3}}
\nc{\PL}[3]     {\JL{Phys. Lett.}{#1}{#2}{#3}}
\nc{\PLB}[3]    {\JL{Phys. Lett.~B}{#1}{#2}{#3}}
\nc{\PTP}[3]    {\JL{Prog. Theor. Phys.}{#1}{#2}{#3}}
\nc{\PTPS}[3]   {\JL{Prog. Theor. Phys. Suppl.}{#1}{#2}{#3}}
\nc{\PTEP}[3]   {\JL{Prog. Theor. Exp. Phys.}{#1}{#2}{#3}}
\nc{\PRep}[3]   {\JL{Phys. Rep.}{#1}{#2}{#3}}
\nc{\PPNP}[3]   {\JL{Prog.\ Part.\ Nucl.\ Phys.}{#1}{#2}{#3}}
\nc{\JPG}[3]     {\JL{J. of Phys. G}{#1}{#2}{#3}}
\nc{\andvol}[3] {{\it ibid.}\JL{}{#1}{#2}{#3}}

\end{document}